\documentclass{article}
\usepackage{spconf,amsmath,graphicx}
\usepackage{multirow}
\usepackage{subcaption}

\title{ENHANCING HEALTHCARE WITH EOG: A NOVEL APPROACH TO SLEEP STAGE CLASSIFICATION}
\name{Suvadeep Maiti, Shivam Kumar Sharma, Raju S. Bapi}
\address{Cognitive Science Lab, International Institute of Information Technology, Hyderabad, India}
\begin{document}
%\ninept
%
\maketitle
\begin{abstract}
We introduce an innovative approach to automated sleep stage classification using EOG signals, addressing the discomfort and impracticality associated with EEG data acquisition. In addition, it is important to note that this approach is untapped in the field, highlighting its potential for novel insights and contributions. Our proposed SE-Resnet-Transformer model provides an accurate classification of five distinct sleep stages from raw EOG signal. Extensive validation on publically available databases (SleepEDF-20, SleepEDF-78, and SHHS) reveals noteworthy performance, with macro-F1 scores of 74.72, 70.63, and 69.26, respectively.  Our model excels in identifying REM sleep, a crucial aspect of sleep disorder investigations. We also provide insight into the internal mechanisms of our model using techniques such as 1D-GradCAM and t-SNE plots. Our method improves the accessibility of sleep stage classification while decreasing the need for EEG modalities. This development will have promising implications for healthcare and the incorporation of wearable technology into sleep studies, thereby advancing the field's potential for enhanced diagnostics and patient comfort. 

\end{abstract}
\section{Introduction}
\label{sec:intro}

The importance of sleep for overall well-being cannot be overstated, and the monitoring of sleep stages is crucial for maintaining optimal health. Sleep is classified into five stages by the American Academy of Sleep Medicine (AASM) \cite{AASM}. Historically, the conventional approach has entailed the utilization of labor-intensive polysomnography (PSG) which incorporates the measurement of electroencephalography (EEG), electrooculography (EOG), and electromyography (EMG) data. The utilization of machine learning and deep learning holds significant potential in the automation of sleep-stage classification, whereby EEG is frequently employed as a prevalent modality. Nevertheless, the process of EEG acquisition has the potential to disturb normal sleep patterns. Therefore, the identification of less invasive yet precise methods is crucial for enhancing the classification of sleep stages. Therefore, the decision has been made to focus our efforts on EOG signals as a viable alternative. 

EOG recordings are utilized to measure eye activity, which serves as a crucial indicator for differentiating between non-rapid eye movement (NREM) and rapid eye movement (REM) sleep stages. EOG signals possess notable advantages owing to their capacity for little sleep disruption and comparatively uncomplicated electrode positioning. Furthermore, it has been observed that EOG signals often display contamination originating from EEG signals, implying a potential association between these two types of signals \cite{2,3}. 

Only a limited number of studies have delved into the realm of EOG-based sleep stage classification, with \cite{EOGNET} focusing exclusively on single-channel EOG data. On the other hand, previous research conducted by \cite{deepsleepnet, xsleepnet,eegbased} has mostly focused on the analysis of EEG data. The objective of our research is to develop a model that performs well in classifying sleep stages using single-channel  EOG alone. The utility of EOG extends beyond the classification of sleep stages, as it shows potential for detecting REM sleep. This is a crucial aspect in the diagnosis of disorders such as REM Sleep Behaviour Disorder (RBD) \cite{remdisorder, remsleep}, Narcolepsy \cite{narcolepsy}, Nightmares, Night Terrors, and Sleep-Related Eating Disorder (SRED), as well as other parasomnias. Moreover, such proof-of-concept can pave the way for future wearable solutions for sleep quality assessment.

Our contributions can be summarized as follows: 

\begin{itemize}
\item We introduce a SEResnet~\cite{seresnet}-Transformer \cite{transformer}-based model 
   designed to enhance the classification of sleep stages using single-channel EOG data tested on multiple publicly available datasets such as SleepEDF-20, SleepEDF-78~\cite{sleepedf}, and SHHS~\cite{shhs}. The proposed model consistently outperforms the existing models. 

\item Our model demonstrates exceptional performance in the identification of the REM stage, crucial for assessing sleep disorders. Transparency is improved through the utilization of 1D-GradCAM~\cite{gradcam} and t-SNE plots, which effectively highlight the mechanisms employed by our method in accurately classifying REM stages.

% \item A notable achievement of our model is its exceptional 
%    proficiency in detecting the REM stage, a critical factor in analyzing sleep disorders. 

% \item We additionally provide insights into our model's inner workings by using 1D-GradCAM and tSNE plots.
\end{itemize}

\begin{figure*}[htbp]
  \centering
  \begin{minipage}[t]{0.70\linewidth} % Adjust the width as needed
    \centering
    \includegraphics[width=0.8\linewidth]{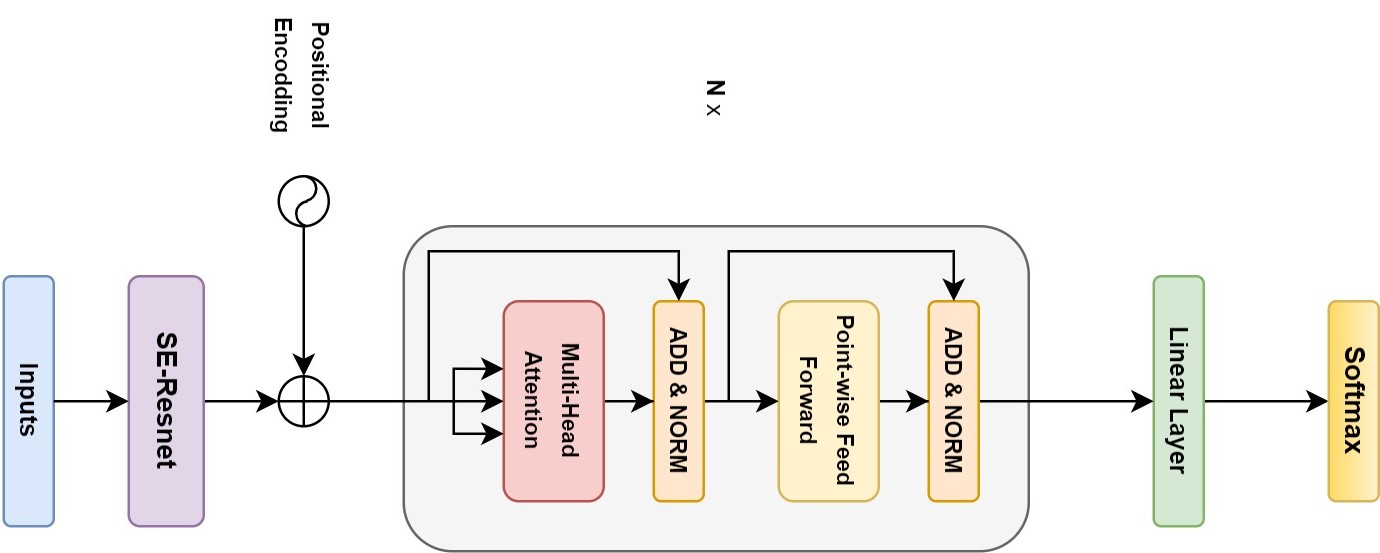} % Adjust the width as needed
    \caption{Proposed Model Architecture}
    \label{fig:image1}
  \end{minipage}
  % \hfill
  \begin{minipage}[t]{0.25\linewidth} % Adjust the width as needed
    \centering
    \includegraphics[width=0.7\linewidth]{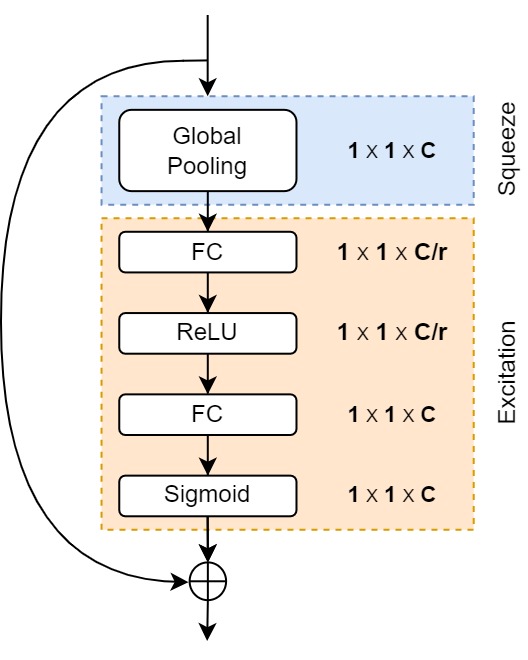} % Adjust the width as needed
    \caption{SE-Block}
    \label{fig:image2}
  \end{minipage}
\end{figure*}

\section{METHODOLOGY}
\subsection{Contextual Input}

Previous research classified sleep stages using a 30-second period of EEG signal (considered as 1 epoch). According to standard practices, sleep stage classification relies on the present epoch as well as the surrounding ones too. In the proposed method that uses EOG alone, we create contextual input from the present and nearby epochs as well. So we used contiguous EOG epochs which we designate as a \emph{window}.

\subsection{SE-Resnet}

The proposed architecture is shown in Fig-\ref{fig:image1}, comprising SE-Resnet  that incorporates a Squeeze-and-Excitation (SE) block (in Fig-\ref{fig:image2}) in each residual block. The SE block enhances network's capability by dynamically recalibrating feature maps. Each residual block normally has two or three convolutional layers, followed by a SE block and shortcut connection. This block handles two crucial operations: ``squeeze" and ``excitation." The ``squeeze" stage gathers global data from all channels to create a channel descriptor vector. The vector is used to selectively reweight channels during the "excitation" process. This approach prioritizes essential channels and suppresses less important ones, enhancing the network's feature representation. 

The reasoning for doing feature extraction is in the existence of unwanted noise or extraneous components within the raw EOG data, which may hinder the accurate classification process. In sleep stage classification, feature extraction is crucial for accurate predictions, features extracted from unprocessed EOG data attempts to simplify and condense significant information. This phase includes carefully selecting characteristics that accurately portray sleep stage-specific patterns while minimizing noise. 

\subsection{Transformer}

The Transformer is a neural network model architecture that has had a significant impact on the field of natural language processing. The underlying principle of transformer is rooted in a mechanism referred to as "self-attention," which is recognized for its ability to parallelize computations and effectively process sequential input. Transformers have gained significant prominence in various applications such as machine translation, text synthesis, and other related activities. They typically comprise an encoder-decoder architecture, or alternatively, a standalone encoder for specific purposes such as text classification. In pur approach we are using juts the encoder part of the transformer. 

EOG signals possess an inherent sequential nature and display intricate temporal interdependencies. Transformers demonstrate exceptional proficiency in modeling sequential data due to their utilization of self-attention mechanisms, which effectively record intricate correlations among distinct time steps within the data. This skill is essential for distinguishing patterns in EOG signals that correlate to particular sleep stages.

\subsection{Classifier}

The architectural design of the model concludes in a linear layer that receives input from the transformer module. This final layer has softmax activation that gives probability distribution over the number of classes.

\section{EXPERIMENTS}
\label{sec:pagestyle}

\begin{table*}[]
\centering
\caption{Classification performance of our model on the SleepEDF-20, SleepEDF-78 and SHHS datasets with single channel raw-EOG input}
\label{tab:TAB1}
\begin{tabular}{|c|ccc|ccccc|}
\hline
\multirow{2}{*}{\textbf{Dataset}} & \multicolumn{3}{c|}{\textbf{Overall Metrices}}                                     & \multicolumn{5}{c|}{\textbf{Per class F1-score}}                                                                                                        \\ \cline{2-9} 
                                  & \multicolumn{1}{c|}{\textbf{ACC}} & \multicolumn{1}{c|}{\textbf{MF1}} & \textbf{k} & \multicolumn{1}{c|}{\textbf{W}} & \multicolumn{1}{c|}{\textbf{N1}} & \multicolumn{1}{c|}{\textbf{N2}} & \multicolumn{1}{c|}{\textbf{N3}} & \textbf{REM} \\ \hline
\textbf{SleepEDF-20}              & \multicolumn{1}{c|}{79.26}        & \multicolumn{1}{c|}{74.72}        & 0.72       & \multicolumn{1}{c|}{81.57}      & \multicolumn{1}{c|}{45.98}       & \multicolumn{1}{c|}{86.36}       & \multicolumn{1}{c|}{81.57}       & 78.13        \\ \hline
\textbf{SlepEDF-78}               & \multicolumn{1}{c|}{73.56}        & \multicolumn{1}{c|}{70.63}        & 0.68       & \multicolumn{1}{c|}{82.80}      & \multicolumn{1}{c|}{31.81}       & \multicolumn{1}{c|}{81.49}       & \multicolumn{1}{c|}{79.75}       & 77.29        \\ \hline
\textbf{SHHS}                     & \multicolumn{1}{c|}{79.04}        & \multicolumn{1}{c|}{69.26}        & 0.71       & \multicolumn{1}{c|}{83.44}      & \multicolumn{1}{c|}{24.1}        & \multicolumn{1}{c|}{82.81}       & \multicolumn{1}{c|}{74.95}       & 80.97        \\ \hline
\end{tabular}
\end{table*}

\begin{figure*}
	\centering
		\includegraphics[scale=.80]{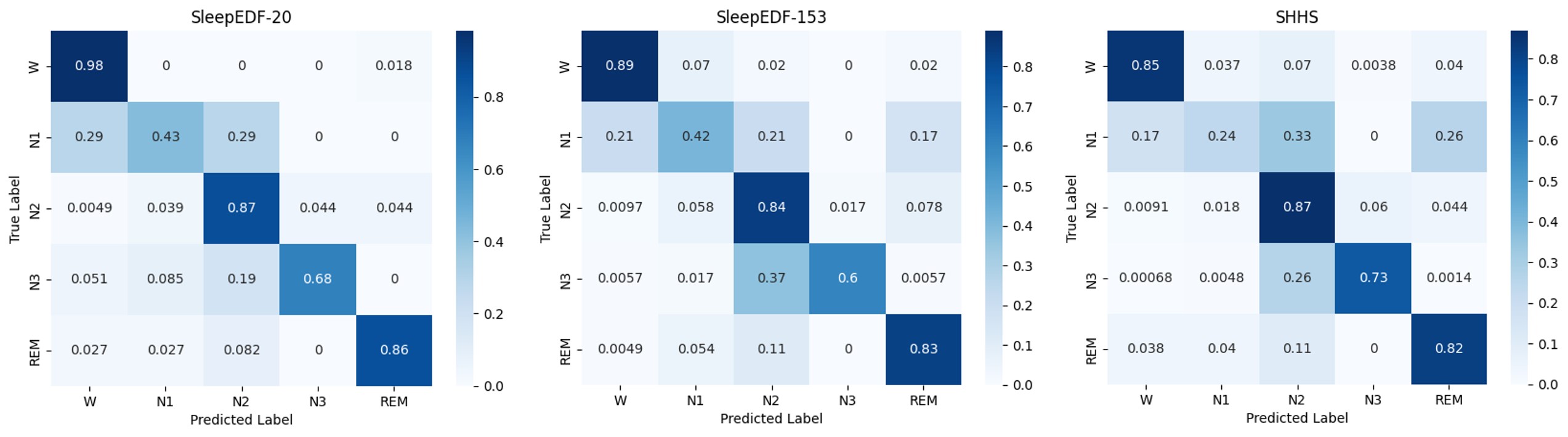}
	\caption{Normalized Confusion matrices of Fold-0 on SleepEDF-20, SleepEDF-78 and SHHS dataset}
	\label{FIG:3}
\end{figure*}

\subsection{Data and Preprocessing}
In our study, we conducted an evaluation of our deep learning model using raw EOG data obtained from three publicly available datasets: SleepEDF-20, SleepEDF-78, and SHHS (Sleep Heart Health Study). These datasets are widely recognized in the field of sleep research and provide valuable resources for understanding sleep patterns and stages. The SleepEDF-20 dataset consists of 39 PSG records from 20 healthy individuals aged 25-34. Manual sleep stage classification based on the Rechtschaffen and Kales (R\&K) standard criteria is available, allowing us to assess sleep stages such as Wake (W), N1, N2, N3 (combining N3 and N4), REM, and others. We excluded the Movement (M) and UNKNOWN categories for our analysis. The SleepEDF-78 dataset is an expanded version of SleepEDF-20, offering PSG recordings from 78 participants. SHHS dataset comprises 6,441 subjects, each with a single full-night PSG recording. Sleep scoring was performed using AASM criteria, with N3 and N4 stages merged into N3. We selected 329 participants with regular sleep patterns for our analysis.

\subsection{Experimental setup}
We utilized the Adam optimizer with a fixed learning rate of 0.001 during the optimization process. This adaptive algorithm combines momentum and squared gradients to update model parameters effectively. To improve sleep data classification and optimize performance during training, we employed the negative log-likelihood loss. To make efficient use of computational resources, we chose a batch size of 128. For a fair comparison, we conducted 20-fold cross-validation for SleepEDF-20, 10-fold for SleepEDF-78, and 5-fold for the SHHS dataset, ensuring robust evaluation across diverse datasets. 

% \usepackage{multirow}
% \usepackage{multirow}

% \usepackage{multirow}
% \usepackage{multirow}
% \usepackage{multirow}

\subsection{Experimental results}
In Table \ref{tab:TAB1}, we have provided an overview of our model's classification performance across three different datasets. In Fig.\ref{FIG:3}, we present normalized confusion matrices for the proposed model, specifically using a window size of 9 and a stride of 1, applied to the EOG channel data from SleepEDF-20, SleepEDF-78, and the SHHS dataset. Our evaluation primarily focuses on Macro-averaged F1 scores (MF1), Accuracy (ACC), Cohen's kappa score (k), and F1 scores for individual sleep stages, including W, N1, N2, N3, and REM. The F1 score is a widely accepted metric for assessing model performance in the context of imbalanced datasets, as it has been commonly used in prior research on sleep staging. Each column within the confusion matrices represents the predicted sleep stages, while each row corresponds to the sleep stage annotations provided by expert human annotators. The darker shades along the diagonal of these matrices indicate better classification performance for the respective sleep stage categories. Based on the information presented in Table \ref{tab:TAB1}, it becomes evident that the metrics associated with the N1 sleep stage exhibit lower values compared to the other four stages, namely W, N2, N3, and REM phases. This difference in performance may be attributed to the relatively limited representation of the N1 stage within our dataset when compared to the other sleep stages.

Table \ref{tab:TAB2} presents a summary of the comparison results obtained from the SleepEDF-20 dataset. It is evident that models trained using EEG data consistently outperform those trained using EOG data, aligning with previous studies. This underscores the significance of EEG components in accurately identifying sleep stages. Our proposed method outperformed all other EOG-trained models, achieving the highest ACC (79.26\%), F1 score (74.72), and k (0.72) for single-channel EOG. For single-channel EEG, it yielded impressive results with an ACC of 82.42\%, an F1-score of 76.54, and a k of 0.75. Our model's performance on EOG channel is on par with models trained using EEG. This suggests that EOG has the potential to be a valuable modality for sleep staging. The achieved accuracy of 79.26\% meets the requirements for various applications, including community health care, home-based sleep monitoring, and even clinical settings. 

\begin{table}[]
\centering
\caption{Result comparison with different methods on SleepEDF-20 dataset. Our model's performance on the EOG signal is shown in bold.}
\label{tab:TAB2}
\begin{tabular}{ccccc}
\hline
\multirow{2}{*}{\textbf{Methods}} & \multirow{2}{*}{\textbf{Input}} & \multicolumn{3}{c}{\textbf{Overall results}}    \\ \cline{3-5} 
                                  &                                 & \textbf{ACC}   & \textbf{F1}    & \textbf{k}    \\
\multirow{2}{*}{DeepSleepNet}\cite{deepsleepnet}     & EEG                             & 82.0             & 76.9             & 0.76            \\
                                  & EOG                             & 73.5             & 68.2             & 0.66            \\
\multirow{2}{*}{EOGNet}\cite{EOGNET}           & EEG                             & 79.6           & 73.2           & 0.71         \\
                                  & EOG                             & 76.3           & 69.3           & 0.67          \\
\multirow{2}{*}{Our method}       & EEG                             & 82.4            & 76.5             & 0.75            \\
                                  & \textbf{EOG}                    & \textbf{79.3} & \textbf{74.7} & \textbf{0.72} \\ \hline
\end{tabular}
\end{table}
\subsection{Ablation studies}
Systematic ablation studies were conducted on the SleepEDF-20 dataset using 20 different folds. The study involved a thorough examination of how various components and features affected our model's performance. We experimented with different window widths and strides to optimize the model's F1 score, discovering that a window size of 9 and a stride of 1 yielded the best results. This larger window size captured a broader temporal context, potentially extracting more valuable information from EOG signals. The model was trained with a window size of 9 and a stride of 4, and subsequently evaluated using the same window size but with a stride of 1, effectively utilizing only 25\% of the dataset for training. This training strategy demonstrated comparable performance to training with the entire dataset but exhibited a remarkable eightfold improvement in training speed.

\section{MODEL INTERPRETABILITY}
\label{sec:typestyle}

\subsection{t-SNE}
\begin{figure}
	\centering
		\includegraphics[scale=.27]{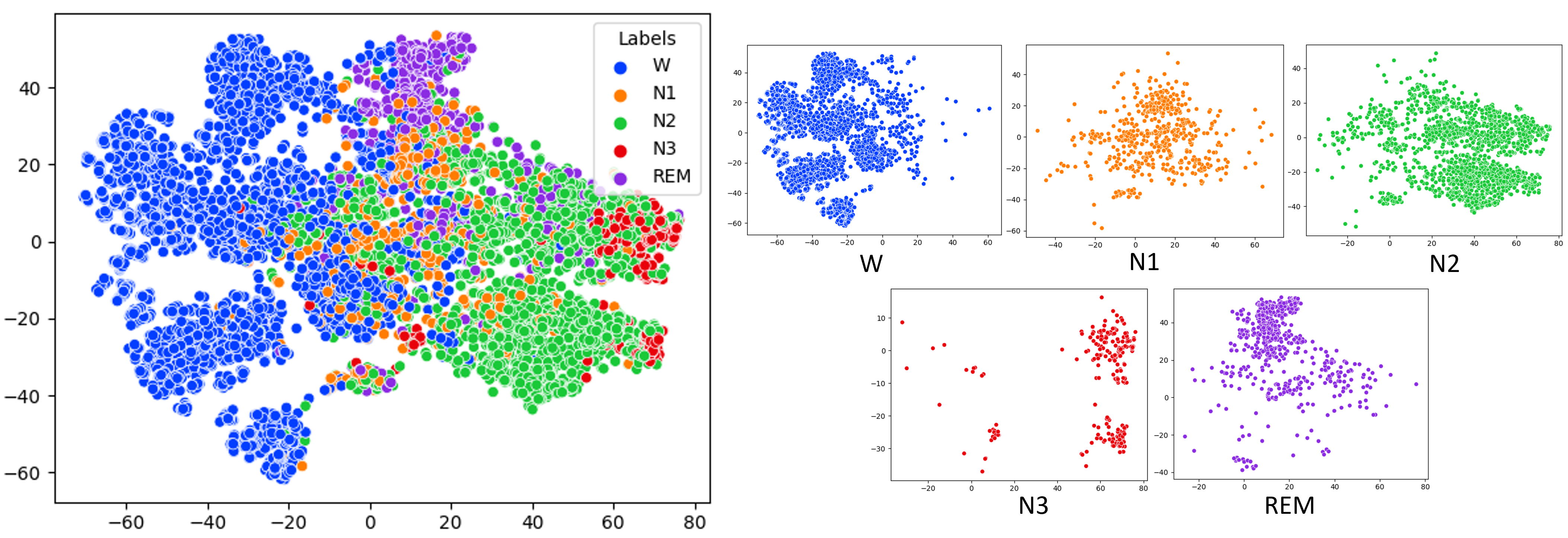}
	\caption{t-SNE plot of extracted features from the SE-Resnet's last convolutional layer.
}
	\label{FIG:4}
\end{figure}

In this study, we illustrated the complex features that were extracted from the final convolutional layer of the SE-Resnet feature extraction module using the t-Distributed Stochastic Neighbour Embedding (t-SNE) technique. The t-SNE visualization (in Fig \ref{FIG:4}) revealed a clustering pattern that corresponded closely with the existence of five distinct classes. This result demonstrates the model's ability to extract essential features from raw EOG data.
It is interesting that, apart from the N1 class, all other classes demonstrated clearly defined and distinct boundaries. Significantly, a considerable level of overlap was seen for N1 class with W, N2 and REM classes. The finding implies that the model encounters difficulties in reliably differentiating instances within the N1 category.

% This claim is additionally substantiated by cases in which the N1 sleep stage was misclassified as either the REM or W sleep stages, as demonstrated in the relevant confusion matrices depicted in Figure-\ref{FIG:3}.

\begin{figure}
	\centering
		\includegraphics[scale=.28]{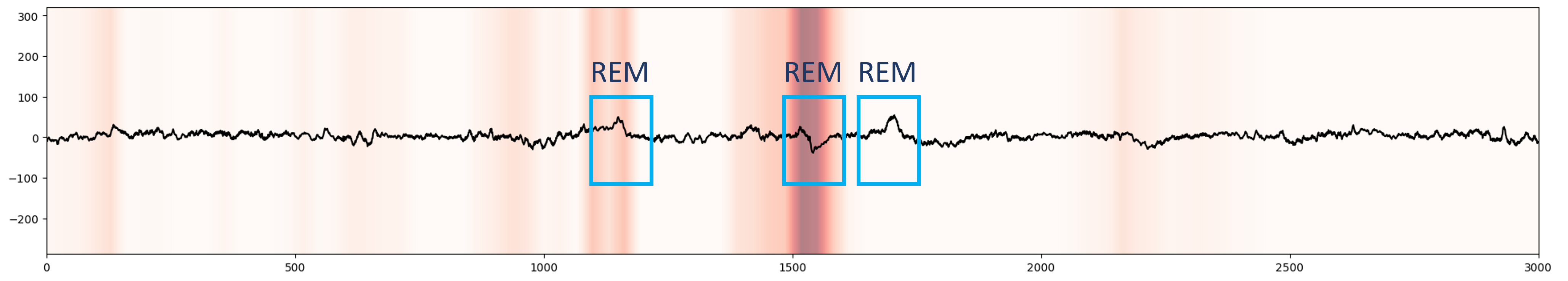}
	\caption{1D-GradCAM visualization of raw EOG epochs and boxes in blue indicating rapid eye movement during REM sleep stage.}
	\label{FIG:5}
\end{figure}

\subsection{GradCAM}

Fig \ref{FIG:5}, displays a one-dimensional GradCAM visualization that represents a rapid eye movement (REM) sleep epoch. The visualization employs a range of red hues to emphasize particular segments of the input signal that have a major influence on the model's final categorization decision. The picture also includes blue boxes that serve to identify the presence of rapid eye movement artifacts, as identified by a sleep expert. Our model exhibits notable effectiveness in accurately identifying REM sleep in the raw EOG signal. The result shows that the model is able to prioritize the same segments of signal that experts use to classify sleep epochs accurately.

\section{DISCUSSION}
In summary, our study presents a novel SeResnet-Transformer model that has been developed for the accurate classification of sleep stages using single-channel EOG data. This model provides a viable alternative to conventional EEG-based approaches. Although our model frequently demonstrates superior performance compared to existing approaches, it notably excels in recognizing the REM sleep stage. However, we recognize the necessity for further enhancements in the detection of the N1 stage, as its representation in our datasets is quite restricted. In the future, our objective is to modify our model to facilitate signal gathering using wearable devices that are convenient, such as eye masks and spectacles. This adaptation will enhance the applicability of our model in long-term sleep monitoring scenarios. Furthermore, our research aims to investigate the incorporation of cardiorespiratory and movement signals with EOG data, capitalizing on their potential for improving the precision and practicality of our approach in the classification of sleep stages.

\bibliographystyle{IEEEbib}
\bibliography{strings,refs}

\end{document}